# On the Achievable Rates of the Diamond Relay Channel with Conferencing Links

Chuan Huang, *Student Member, IEEE,* Jinhua Jiang, *Member, IEEE,*
Shuguang Cui, *Member, IEEE,*

**Abstract**

We consider a half-duplex diamond relay channel, which consists of one source-destination pair and two relay nodes connected with two-way rate-limited out-of-band conferencing links. Three basic schemes and their achievable rates are studied: For the decode-and-forward (DF) scheme, we obtain the achievable rate by letting the source send a common message and two private messages; for the compress-and-forward (CF) scheme, we exploit the conferencing links to help with the compression of the received signals, or to exchange messages intended for the second hop to introduce certain cooperation; for the amplify-and-forward (AF) scheme, we study the optimal combining strategy between the received signals from the source and the conferencing link. Moreover, we show that these schemes could achieve the capacity upper bound under certain conditions. Finally, we evaluate the various rates for the Gaussian case with numerical results.

## I. Introduction

In most beyond-3G wireless technologies such as WiMAX and 3GPP UMTS Long Term Evolution (LTE), the concept of relaying is introduced to provide coverage extension and increase

Chuan Huang and Shuguang Cui are with the Department of Electrical and Computer Engineering, Texas A&M University, College Station, TX, 77843. Emails: {huangch, cui}@tamu.edu.

Jinhua Jiang is with the Department of Electrical Engineering, Stanford University, Stanford, CA, 94305, Email: jhjiang@stanford.edu.

capacity. From the information-theoretical viewpoint, the capacity bounds of the traditional three-node relay channel have been well studied [1]–[4], and various achievable schemes, such as decode-and-forward (DF) and compress-and-forward (CF), have been proposed. For the half-duplex relay channel, in [4] and the references therein the authors have studied the achievable rates and the power allocation problem.

For the case with two relay nodes and no direct link between the source and the destination, termed as the diamond relay channel, various achievable rates were studied in [5]–[8]. In particular, the authors in [5] discussed the capacity upper bound and the achievable rates using the DF and amplify-and-forward (AF) schemes under the full-duplex relaying mode. Under the half-duplex mode, the authors in [6] discussed the achievable rates using two time-sharing schemes, i.e., the simultaneous relaying and alternative relaying schemes. By further exploring partial collaboration between the two relays, the authors in [7], [8] developed some DF schemes based on dirty paper coding (DPC) and block Markov encoding (BME), where the DF scheme is shown to be optimal in some special cases [7].

In practical communication systems, some nodes might have extra out-of-band connections with the others, e.g., through blue-tooth, WiFi, optical fiber, etc., to exchange certain information and improve the overall system performance. From the information-theoretical viewpoint, such kind of interaction can be modeled as nodes conferencing [9]–[13]. Specifically, for multiple access channel (MAC) [9], encoder conferencing was used to exchange part of the source messages, and it is proved that one-round conferencing scheme is optimal. For the broadcast channel (BC) in [10], the decoders was designed to first compress the received signal, and then transmit the corresponding binning index number to the other through the receivers conferencing links. In [10], [11], it was shown that the one-round scheme is optimal for physically degraded BC channel, while the two-round and three-round schemes can outperform the one-round one in general cases. Moreover, in [12] and [13], the achievable rate of the compound MAC channel with transmitter and receiver conferencing was discussed, and some capacity results for the degraded cases were provided.

In this paper, we consider a two-hop diamond relay channel, which contains two half-duplex relay nodes. We assume that the relays can conduct conferencing with each other via some orthogonal out-of-band links [14]. Generally, the conferencing links can be used to exchange a compressed version of the received signals at the relays [10], part of the messages intended to the destination between the two relays [9], or just to forward the received signal to the other relay [6]. With these ideas, we develop relaying schemes based on the DF, CF, and AF schemes by exploiting the inter-relay conferencing, for both the cases of discrete memoryless channel (DMC) and Gaussian channel. Moreover, in stead of considering multi-round conferencing scheme [10], [11], we just concentrate on the simple one-round conferencing scheme, which means that the relays simultaneously process their received signal and conduct conferencing with the other in the same time slot. The main results of the paper are summarized as follows:

1) For the DF relaying scheme, we let the source to transmit one common message to both relays and one private message to each relay. We prove that for the DMC case, the DF scheme could achieve the capacity cut-set bound just with finite conferencing link rates; for the Gaussian case, the cut-set bound is asymptotically achieved when the source-to-relay link signal-to-noise ratios (SNR) go to infinity.
2) For the CF relaying scheme, we develop three schemes: one using conferencing links to help the compression, and the other two using them to partially or fully exchange the binning index of the compressed receiver signal. We prove that for the Gaussian case, when the SNRs of the BC channel or the MAC channel go to infinity, the capacity upper bound is asymptotically achievable.
3) For the AF relaying scheme, we investigate the optimal combining problem between the received signals from the source and the other relay. Generally, it is not a concave problem, while semidefinite relaxation can be applied to transform it to a quasi-concave problem.

The remainder of the paper is organized as follows. In Section II, we introduce all the assumptions and channel models. In Section III, we derive the capacity upper bound and the achievable rates for the DF, CF, and AF schemes. Moreover, we discuss some capacity achieving cases. In Section IV, we show some simulation and numerical results. Finally, the paper is concluded in Section VI.

We define the following notations used throughout this paper: $\log(x)$ is the base-2 logarithm; $\text{Tr}(\boldsymbol{A})$ is the trace of matrix $\boldsymbol{A}$; and $\Re(x)$ is real part of $x$.

## II. Assumptions and System Model

In this paper, we consider a diamond relay channel with out-of-band conferencing links between the relays, as shown in Fig. 1, which contains one source node, one destination node, and two relays. There is no direct link between the source and destination. The relay nodes work in a half-duplex mode: the source transmits and the two relays listen in the first time slot; the relays simutaneously transmit and the destination listens in the second time slot. Denote the time fraction allocated to the first slot as $\lambda$, with $\lambda \in (0, 1)$, and the time fraction for the second slot as $\overline{\lambda} = 1 - \lambda$. The capacity of the conferencing link from relay 1 to relay 2 is given $C_{12}$, and $C_{21}$ is defined similarly. Furthermore, these two conferencing links are orthogonal to each other and outside the bandwidth used by the source-to-relay and relay-to-destination links. The time scheduling of the transmissions at the source, relays, and conferencing links is shown in Fig. 2(a) and Fig. 2(b).

In this paper, we assume that for the DF and CF relaying schemes, we adopt the CF scheme as the conferencing strategy; and for the AF relaying scheme, we adopt the AF scheme for conferencing. Due to this assumption, we note that the transmission scheduling schemes for DF, CF, and AF are different: For the DF and CF relaying schemes, the block length of the conferencing link codewords is equal to the sum of those for the source and relay transmission codewords; on the other hand, for the AF relaying scheme, the block lengthes of these three codewords should be the same, and the conferencing link rate is subject to a one-half half-duplex penalty. Moreover, due to the relay conferencing, there will be a one-block delay between the transmissions at the source and the relays, as shown in Fig. 2(a) and Fig. 2(b), which requires the relays to buffer one block of source signals for each relaying operation. Assume that during each block, the communication rate is $R$, and we need to transmit $B$ blocks in total. Thus, the average information rate is $R\frac{B}{B+1} \to R$, as $B$ goes to infinity. In this paper, we focus on one-block transmission and the associated coding scheme without specifying the delay in the proof of the achievability.

For the Gaussian case, we further define the following channel input-output relationship. The received signal $y_i$ from the source at the $i$-th relay ($i = 1, 2$) is given as

$$y_i = h_i x + n_i, \ i = 1, 2, \tag{1}$$

where $x$ is the signal transmitted by the source with power $P_S$, $h_i$ is the complex channel gain of the $i$-th source-to-relay link, and $n_i$'s are the independently and identically distributed (i.i.d.) circularly symmetric complex Gaussian (CSCG) noise with distribution $\mathcal{CN}(0, 1)$.

In the second hop, signal $x_i$ with average power $P_R$, is transmitted from the $i$-th relay to the destination; and the received signal $y$ at the destination is given as

$$y = \sum_{i=1}^{2} g_i x_i + n, \tag{2}$$

where $g_i$ is the complex channel gain of the $i$-th relay-to-destination link, and $n$ is the CSCG noise with distribution $\mathcal{CN}(0,1)$. For convenience, we define the link SNRs as

$$\gamma_i = |h_i|^2 P_S,\ \widetilde{\gamma}_i = |g_i|^2 P_R,\ i = 1, 2. \tag{3}$$

## III. Capacity Upper Bound and Achievable Rates

In this section, we exam the capacity upper bound and the achievable rates of the considered channel with the following three relaying schemes: DF, CF, and AF, respectively. Moreover, we prove some capacity achieving results under special conditions. To be concise, in each relaying scheme we generically describe the coding scheme for the $i$-th relay ($i = 1, 2$), where we use $(3 - i)$ to refer to the other relay index for the convenience of description.

### *A. Capacity Upper Bound*

In this subsection, we first study the capacity upper bound for the considered channel. The upper bound is derived by the cut-set theory [1].

*Theorem 3.1:* The capacity upper bound for the discrete memoryless diamond relay channel with conferencing links is given as

$$C_{\text{upper}} \leq \frac{I\left(X; Y_1, Y_2\right) I\left(X_1, X_2; Y\right)}{I\left(X; Y_1, Y_2\right) + I\left(X_1, X_2; Y\right)}, \tag{4}$$

over distribution $p(x)p(y_1, y_2|x)p(x_1, x_2)p(y|x_1, x_2)$.

*Proof:* By the cut-set bound, we have $C_{\text{upper}} \leq \min\left\{\lambda I\left(X; Y_1, Y_2\right), \overline{\lambda} I\left(X_1, X_2; Y\right)\right\}$, which comes from the broadcast (BC) cut-set and multiple access (MAC) cut-set [6], [20]. We then optimize over $\lambda$ to obtain a better bound, and the minimum value is achieved iff the two terms are equal, which means $\lambda^* = \frac{I(X_1, X_2; Y)}{I(X; Y_1, Y_2) + I(X_1, X_2; Y)}$. With this optimal $\lambda$, we obtain the upper bound in (4). ■

This theorem implies that the capacity upper bound is achieved only when a common message with the rate given in (4) is sent and can be perfectly decoded by both of the relays in a cooperative way.

For the Gaussian case, since these two-hop communications are independent with each other (i.e., maximizing $C_{\text{upper}}$ means maximizing $I(X;Y_1,Y_2)$ and $I(X_1,X_2;Y)$, respectively), we choose $X$, $X_1$, and $X_2$ to be independent CSCG with distribution $\mathcal{CN}(0,P_S)$, $\mathcal{CN}(0,P_R)$, and $\mathcal{CN}(0,P_R)$, respectively; and the corresponding capacity upper bound is given by the following corollary.

*Corollary 3.1:* For the Gaussian case, we have the following capacity upper bound

$$C_{\text{upper}} \le \frac{\log(1+\gamma_1+\gamma_2)\log\left(1+\widetilde{\gamma}_1+\widetilde{\gamma}_2+2\sqrt{\widetilde{\gamma}_1\widetilde{\gamma}_2}\right)}{\log(1+\gamma_1+\gamma_2)+\log\left(1+\widetilde{\gamma}_1+\widetilde{\gamma}_2+2\sqrt{\widetilde{\gamma}_1\widetilde{\gamma}_2}\right)}. \quad (5)$$

### B. DF Achievable Rate

**Main idea:** For the DF scheme, the source transmits three messages: one common message $w_0$ to both of the relays, and one private message to each of the relays, denoted as $w_1$ and $w_2$, respectively. In the $i$-th relay, it compresses the received signal from the source, and sends the corresponding binning index through the conferencing link to the other relay, which helps with decoding the desired common message. In the second hop, the channel is indeed a MAC with common information. In the next, we first consider the DMC case and then consider the Gaussian case.

*1) DF Rate for the DMC Case:* We first focus on the first hop that is a BC channel with receiver one-round conferencing, for which the authors in [10] investigated the two cases with two independent messages and only one common message, respectively. In this subsection, we extend their results with a more general coding scheme, and have the following lemma.

*Lemma 3.1:* The achievable rate region of the general discrete memoryless BC with common message and decoder conferencing is given as

$$R_{\text{BC}} = \bigcup_{p(u_0)p(u_1|u_0)p(u_2|u_0)x(u_0,u_1,u_2)p(y_1,y_2|x)p(\hat{y}_1|y_1)p(\hat{y}_2|y_2)} \left\{ \begin{array}{l} (R_0,R_1,R_2): \; R_0,R_1,R_2 \ge 0, \\ R_0+R_i \le \lambda I\left(U_0,U_i;\hat{Y}_{3-i},Y_i\right), \\ R_0+R_1+R_2 \le \lambda I\left(U_i;\hat{Y}_{3-i},Y_i|U_0\right)+\lambda I\left(U_0,U_{3-i};\hat{Y}_i,Y_{3-i}\right)-\lambda I\left(U_1;U_2|U_0\right), \\ 2R_0+R_1+R_2 \le \lambda I\left(U_0,U_1;\hat{Y}_2,Y_1\right)+\lambda I\left(U_0,U_2;\hat{Y}_1,Y_2\right)-\lambda I\left(U_1;U_2|U_0\right), \end{array} \right\}. \quad (6)$$

subject to the following constraints

$$C_{i,3-i} \geq \lambda I\left(\hat{Y}_i; Y_i\right) - \lambda I\left(\hat{Y}_i; Y_{3-i}\right), \ i = 1, 2, \tag{7}$$

where $R_0$, $R_1$, and $R_2$ are the rates of the common message, the private message for the first relay and the private message for the second relay, respectively, and $U_0$, $U_1$, $U_2$, $\hat{Y}_1$, and $\hat{Y}_2$ are auxiliary random variables defined on arbitrary finite sets with the distribution given in (6).

*Proof:* See Appendix A. ■

For the second hop, i.e., the MAC with common message, the achievable rate region is well studied, which is presented in the following lemma.

*Lemma 3.2:* The achievable rate for discrete memoryless MAC with common message is given as [16]

$$R_{\mathrm{MAC}} = \bigcup_{p(x_0)p(x_1|u)p(x_2|u)p(y|x_1,x_2)} \left\{ \begin{array}{l} (R_0, R_1, R_2): \ R_0, R_1, R_2 \geq 0, \\ R_1 \leq \overline{\lambda} I\left(X_1; Y|U, X_2\right), \\ R_2 \leq \overline{\lambda} I\left(X_2; Y|U, X_1\right), \\ R_1 + R_2 \leq \overline{\lambda} I\left(X_1, X_2; Y|U\right), \\ R_0 + R_1 + R_2 \leq \overline{\lambda} I\left(U, X_1, X_2; Y\right). \end{array} \right\}, \tag{8}$$

where $U$ is an auxiliary random variable defined on arbitrary finite set with the distribution given in (8).

From the Lemmas 3.1 and 3.2, we have the following theorem for the achievable rate of the considered diamond relay channel.

*Theorem 3.2:* The achievable rate of the DMC diamond relay channel with conferencing links is given as

$$R_{\mathrm{DF}} = \max_{\lambda, (R_0, R_1, R_2) \in R_{\mathrm{BC}} \bigcap R_{\mathrm{MAC}}} R_0 + R_1 + R_2. \tag{9}$$

*Corollary 3.2:* For the DMC case, the capacity upper bound given in (4) is achieved with finite $C_{12}$ and $C_{21}$, which are upper-bounded as

$$\left\{ \begin{array}{l} C_{12} \leq \lambda^* H\left(Y_1|Y_2\right) \\ C_{21} \leq \lambda^* H\left(Y_2|Y_1\right) \end{array} \right., \tag{10}$$

where $\lambda^*$ is defined in Theorem 3.1.

*Proof:* First, we notice that using one common message only is sufficient to achieve the capacity upper bound; so we focus on the case with only one common message transmitted. In (6), by choosing $U_1$ and $U_2$ as constants (also by Theorem 3 in [10]), we obtain

$$R \leq \lambda^* \min\left\{ I\left(X; Y_1, \hat{Y}_2\right), I\left(X; Y_2, \hat{Y}_1\right)\right\}, \tag{11}$$

subject to $C_{12} \geq \lambda^* I\left(\hat{Y}_1; Y_1\right) - \lambda^* I\left(\hat{Y}_1; Y_2\right)$ and $C_{21} \geq \lambda^* I\left(\hat{Y}_2; Y_2\right) - \lambda^* I\left(\hat{Y}_2; Y_1\right)$. We choose $\hat{Y}_1 = Y_1$ and $\hat{Y}_2 = Y_2$, and obtain (10). ■

*Remark 3.1:* This corollary only gives a maximum value for $C_{i,3-i}$ to achieve the capacity upper bound, and the upper bounds of $C_{i,3-i}$, $i = 1, 2$, can be regarded as the maximum difference between the two received signals at the relays. In the proof, we point out that this result is only a sufficient condition, and this is due to the fact that the cut-set bound is relatively loose under general channel conditions [17]. Another reason is that for some cases, the DF scheme can achieve with capacity upper bound without conferencing. For example, when the BC channel part is deterministic, i.e., $Y_1 = f_1(X)$ and $Y_2 = f_2(X)$, where $f_1$ and $f_2$ are some deterministic functions, the BC cut-set bound is achieved by sending one private message to each relay [18], and this means that conferencing will not introduce any improvement.

*Remark 3.2:* With $C_{i,3-i} = 0$, we claim that our proposed scheme is equivalent to the traditional DF scheme without conferencing. For such a case, we choose $\hat{Y}_1$ and $\hat{Y}_2$ as constants, and $R_{\text{BC}}$ will degrade to the rate region of a BC channel with common message. Moreover, for the Gaussian BC channel, we only need to transmit one common message to both relays and one private message to the better relay [6]. Thus, our scheme is a generalization of the traditional DF scheme, and our DF rate will be the same as or higher than that without conferencing.

*2) DF Rate for the Gaussian Case:* First, we consider the BC part. The first hop is indeed a vector BC with correlated noises, which is not physically degraded in general. Therefore, it is possible to transmit a unique private message to each relay. For the compression at the relays, we choose $\hat{Y}_i = Y_i + N_{i,3-i}$, where $N_{i,3-i}$ is a CSCG random variable distributed as $\mathcal{CN}\left(0, \sigma^2_{i,3-i}\right)$. It is easy to check that the Pareto boundary of the rate region over $(R_0, R_1, R_2)$ is achieved when the variances of the compression noises are minimized, which means that the equality in (7) is achieved, i.e., the compression noise is set to have

$$\sigma^2_{i,3-i} = \frac{1 + \gamma_1 + \gamma_2}{\left(\gamma_{3-i} + 1\right)\left(2^{C_{i,3-i}/\lambda} - 1\right)}. \tag{12}$$

We now discuss the coding scheme for the Gaussian BC, which combines DPC and superposition coding [19]. We choose the transmitting signal $X = X_0 + X_1 + X_2$, where $X_0$, $X_1$, and $X_2$ denote the common message and the private messages intended to relay 1 and relay 2, respectively, and they are independent zero mean CSCG random variables with variances $\overline{\mu} P_S$, $\mu_1 P_S$, and $\mu_2 P_S$, respectively, where the positive parameters $\overline{\mu}$, $\mu_1$, and $\mu_2$ are power allocation factors for $X_0$, $X_1$, and $X_2$, respectively, with $\overline{\mu} + \mu_1 + \mu_2 = 1$.

At the relays, the common message is first decoded by both of them, and then each relay decodes its intended private message. Private messages are encoded using DPC [19]: If we first encode $X_1$, we use $X_1$ as a state information to help with encoding $X_2$; and in the decoding process, relay 2 can decode $X_2$ without interference from $X_1$; on the other hand, we can exchange the encoding and decoding orders to possibly obtain a better rate region. Therefore, the rate region of the first hop is given as

$$R_{\text{BC}} = \text{Conv}\left( \bigcup_{\pi,\mu_1,\mu_2} R\left(\pi, \mu_1, \mu_2\right) \right), \tag{13}$$

where Conv$(\cdot)$ is the convex hull operator, and $R\left(\pi, \mu_1, \mu_2\right)$ is the achievable rate region under a given power allocation scheme $(\mu_1, \mu_2)$ and encoding order $\pi \in \{\pi_{12}, \pi_{21}\}$ with $\pi_{i,3-i}$ meaning that the $i$-th relay's private message is encoded first. Specially, if $X_2$ is encoded first, we have

$$R\left(\pi_{21}, \mu_1, \mu_2\right) = \left\{ \begin{array}{l} (R_0, R_1, R_2) : \\ R_0 \leq \min_{i=1,2} \lambda \log\left(1 + \frac{\overline{\mu}\gamma_1\left(1+\sigma_{21}^2\right)+\overline{\mu}\gamma_2}{((\mu_1+\mu_2)\gamma_1+1)\left(1+\sigma_{21}^2\right)+(\mu_1+\mu_2)\gamma_2}\right) \\ R_1 \leq \lambda \log\left(1 + \mu_1\gamma_1 + \frac{\mu_1\gamma_2}{1+\sigma_{21}^2}\right) \\ R_2 \leq \lambda \log\left(1 + \frac{\mu_2\gamma_2\left(1+\sigma_{12}^2\right)+\mu_2\gamma_1}{(\mu_1\gamma_2+1)\left(1+\sigma_{12}^2\right)+\mu_2\gamma_1}\right) \end{array} \right\}, \tag{14}$$

and $R\left(\pi_{12}, \mu_1, \mu_2\right)$ can be computed similarly.

Next, we consider the MAC part. We choose $X_1 = \sqrt{\overline{\alpha} P} U + \sqrt{\alpha P} V_1$ and $X_2 = \sqrt{\overline{\beta} P} U + \sqrt{\beta P} V_2$, where $U$, $V_1$, and $V_2$ are independent CSCG variables with distribution $\mathcal{CN}\left(0, 1\right)$. Thus, the achievable rate region of the MAC channel with common message is given as

$$\left\{ \begin{array}{l} R_1 \leq \overline{\lambda} \log\left(1 + \alpha\widetilde{\gamma}_1\right) \\ R_2 \leq \overline{\lambda} \log\left(1 + \beta\widetilde{\gamma}_2\right) \\ R_1 + R_2 \leq \overline{\lambda} \log\left(1 + \alpha\widetilde{\gamma}_1 + \beta\widetilde{\gamma}_2\right) \\ R_0 + R_1 + R_2 \leq \overline{\lambda} \log\left(1 + \widetilde{\gamma}_1 + \widetilde{\gamma}_2 + 2\sqrt{\overline{\alpha}\overline{\beta}\widetilde{\gamma}_1\widetilde{\gamma}_2}\right) \end{array} \right. . \tag{15}$$

Therefore, as stated in Theorem 3.2, the DF achievable rate is the maximum sum rate over the intersection of the regions given in (14) and (15).

*Remark 3.3:* From (12), we observe that when $\sigma^2_{i,3-i}$ goes to zero, $C_{i,3-i}$ goes to infinity. In other words, for the Gaussian case, only when $C_{12}$ and $C_{21}$ are infinity, the DF scheme can achieve the capacity upper bound, which is different from the DMC case. Intuitively, for Gaussian channels, the alphabet size of $\mathcal{X}$ is infinite, and each relay cannot reliably decode its counterpart's received signal with the limited help from the other relay.

*Remark 3.4:* When $\gamma_i$ goes to infinity, the optimal $\lambda$ goes to 0, and the capacity upper bound becomes the same as the MAC cut-set bound. In this case, the source only needs to transmit a common message, and both relays can successfully decode it. Therefore, for finite $C_{i,3-i}$ and $\widetilde{\gamma}_i$, the DF scheme can asymptotically achieve the cut-set bound as $\gamma_i$ goes to infinity. On the other hand, when $\gamma_i$ and $C_{i,3-i}$ are fixed, and $\widetilde{\gamma}_i$ goes to infinity, the upper bound cannot be asymptotically achieved. This is due to the fact that the BC cut-set bound cannot be achieved with finite-rate relay conferencing.

### *C. CF Achievable Rates*

In this subsection, we discuss three different coding schemes based on the CF relaying scheme. The first two schemes exploit the conferencing links to partially or completely exchange the binning index of the compressed receiver signals at the relays, and we call them the partial cooperation CF scheme (PCF) and the full cooperation CF scheme (FCF), respectively, which implies how much cooperation is introduced in the MAC part; the third scheme uses the conferencing links to help compression, called as the CCF scheme.

*1) PCF achievable rate:* Here each relay first compresses its received signal as $\hat{Y}_i$ independently and obtains the corresponding binning index. Then, each relay splits the binning index into two sub-messages, and transmit one of them to the other relay by conferencing. In the second hop, the active part of the system is nothing but a MAC channel with a common message. Since we only introduce partially cooperative transmission in the MAC channel, we call it as the partial cooperation CF scheme, i.e., PCF, as defined earlier.

**DMC Case:** We have the following theorem for the achievable rate.

*Theorem 3.3:* The PCF achievable rate for the DMC case is given as

$$R_{\text{PCF}} \leq \max \lambda I \left(X; \hat{Y}_1, \hat{Y}_2\right) \tag{16}$$

$$\text{s. t. } \lambda I \left(\hat{Y}_1; Y_1 | \hat{Y}_2\right) \leq \overline{\lambda} I \left(X_1; Y | U, X_2\right) + C_{12} \tag{17}$$

$$\lambda I \left(\hat{Y}_2; Y_2 | \hat{Y}_1\right) \leq \overline{\lambda} I \left(X_2; Y | U, X_1\right) + C_{21} \tag{18}$$

$$\lambda I \left(\hat{Y}_1, \hat{Y}_2; Y_1, Y_2\right) \leq \min\{\overline{\lambda} I \left(X_1, X_2; Y | U\right) + C_{12} + C_{21}, \overline{\lambda} I \left(X_1, X_2; Y\right)\}, \tag{19}$$

over the distribution $p(x)p(y_1, y_2)p(\hat{y_1}|y_1)p(\hat{y_2}|y_2)p(u)p(x_1|u)\ p(x_2|u)p(y|x_1, x_2)$, and $U$ is an auxiliary random variable similarly defined as before.

The proof of this theorem is trivial: The coding scheme in the first hop is the same as that for the traditional CF scheme in [6]; the second hop with conferencing links is a MAC channel with conferencing encoders and its rate region is given in [9]. By a similar argument to that in [6], we can obtain the PCF rate as shown in this theorem.

*Remark 3.5:* For the case $C_{i,3-i} = 0,\ i = 1, 2$, the PCF scheme is the same as the traditional CF scheme without conferencing [6]; for the case $C_{i,3-i} > 0$, the PCF scheme is not worse than the traditional CF scheme. Note that even when the MAC region is strictly enlarged compared to the case without conferencing, we still cannot claim that the PCF scheme is strictly better than the case without conferencing, since the right-hand side of (19) may not be strictly improved, and when (19) is dominant among these constraints, the PCF rate will be equal to the case without conferencing.

**Gaussian Case:** We define the compression at the relays as $\hat{Y}_i = Y_i + \hat{N}_i,\ i = 1, 2$, where $\hat{N}_i$ is the compression noise with distribution $\mathcal{CN}\left(0, \sigma_i^2\right)$.

*Corollary 3.3:* The PCF achievable rate for the Gaussian case is given as

$$R_{\text{PCF}} = \max_{\lambda,\alpha,\beta,\sigma_1^2,\sigma_2^2} \lambda \log\left(1 + \frac{\gamma_1}{1+\sigma_1^2} + \frac{\gamma_2}{1+\sigma_2^2}\right) \tag{20}$$

$$\text{s. t. } \lambda \log\left(1 + \frac{1}{\sigma_1^2}\left(1 + \frac{\gamma_1\left(1+\sigma_2^2\right)}{1+\sigma_2^2+\gamma_2}\right)\right) \leq \overline{\lambda} \log\left(1 + \alpha\widetilde{\gamma}_1\right) + C_{12}, \tag{21}$$

$$\lambda \log\left(1 + \frac{1}{\sigma_1^2}\left(1 + \frac{\gamma_1\left(1+\sigma_2^2\right)}{1+\sigma_2^2+\gamma_2}\right)\right) \leq \overline{\lambda} \log\left(1 + \beta\widetilde{\gamma}_2\right) + C_{21}, \tag{22}$$

$$\lambda \log\left(1 + \frac{1+\gamma_1}{\sigma_1^2} + \frac{1+\gamma_2}{\sigma_2^2} + \frac{1+\gamma_1+\gamma_2}{\sigma_1^2\sigma_2^2}\right) \leq \min\left\{\overline{\lambda} \log\left(1 + \alpha\widetilde{\gamma}_1 + \beta\widetilde{\gamma}_2\right) + C_{12} + C_{21},\right.$$
$$\left.\overline{\lambda} \log\left(1 + \widetilde{\gamma}_1 + \widetilde{\gamma}_2 + 2\sqrt{\overline{\alpha}\overline{\beta}\widetilde{\gamma}_1\widetilde{\gamma}_2}\right)\right\}. \tag{23}$$

*Remark 3.6:* For given $\gamma_i$, $C_{i,3-i}$, $\alpha$, and $\beta$, when $\widetilde{\gamma}_i \to \infty$, which means that the optimal $\lambda \to 1$, from (21), (22), and (23), we see that both $\sigma_1^2$ and $\sigma_2^2$ scale to 0, and (20) asymptotically achieves the capacity upper bound $\log(1+\gamma_1+\gamma_2)$. Therefore, *when $\widetilde{\gamma}_i \to \infty$, the PCF scheme asymptotically achieve the capacity upper bound*.

*Remark 3.7:* For given $\gamma_i$ and $\widetilde{\gamma}_i$, when $C_{12}$ and $C_{21}$ are large enough, i.e.,

$$C_{12} + C_{21} \geq \log\left(1 + \widetilde{\gamma}_1 + \widetilde{\gamma}_2 + 2\sqrt{\widetilde{\gamma}_1\widetilde{\gamma}_2}\right), \tag{24}$$

the constraints (21) and (22) become redundant, and the CF achievable rate becomes

$$R_{\text{PCF}} = \max_{\lambda,\sigma_1^2,\sigma_2^2} \lambda \log\left(1 + \frac{\gamma_1}{1+\sigma_1^2} + \frac{\gamma_2}{1+\sigma_2^2}\right) \tag{25}$$

$$\text{s. t.} \quad \lambda \log\left(1 + \frac{1+\gamma_1}{\sigma_1^2} + \frac{1+\gamma_2}{\sigma_2^2} + \frac{1+\gamma_1+\gamma_2}{\sigma_1^2\sigma_2^2}\right) \leq \bar{\lambda} \log\left(1 + \widetilde{\gamma}_1 + \widetilde{\gamma}_2 + 2\sqrt{\widetilde{\gamma}_1\widetilde{\gamma}_2}\right). \tag{26}$$

However, since the left-hand side of (26) is strictly larger than $\lambda \log\left(1 + \frac{\gamma_1}{1+\sigma_1^2} + \frac{\gamma_2}{1+\sigma_2^2}\right)$, we cannot find a $\lambda$, which makes (20) equal to the capacity upper bound and satisfies the constraint (26) simultaneously. Thus, *with finite channel gains, the PCF scheme cannot achieve the capacity upper bound even with infinite conferencing rates*.

*Remark 3.8:* For the case that $C_{i,3-i}$ and $\widetilde{\gamma}_i$ are fixed, and $\gamma_i \to \infty$, only if the condition (24) is satisfied, we can approach the capacity upper bound. This is due to the following fact: If we fix $\sigma_1^2$ and $\sigma_2^2$, and choose $\lambda = \frac{\log\left(\widetilde{\gamma}_1+\widetilde{\gamma}_2+2\sqrt{\widetilde{\gamma}_1\widetilde{\gamma}_2}\right)}{\log(1+\gamma_1+\gamma_2)}$, it is easy to check that (20) asymptotically achieves the upper bound, the constraints (21) and (22) become redundant, and (23) asymptotically holds when we have $\gamma_i \to \infty$ and (24) satisfied.

*2) FCF Achievable Rate:* With FCF, after obtaining the compression of the received signal $\hat{Y}_i$, each relay finds the binning index (the number of bins is determined by the corresponding conferencing link rate), and send this binning index to the other relay. Based on its own received signal and the binning index from the other relay, each relay tries to decode the compressed signal of the other relay. Then, we partition the two compressions again into some other bins and transmit the new binning indices to the destination. In this case, each relay has a full knowledge of these two binning indices, and transmits a common message $X_r$ through the MAC channel to the destination. Since we introduce full cooperation over such a MAC channel, we call this scheme as the full cooperation CF scheme, i.e., FCF, as defined earlier.

**DMC Case:** We have the following theorem for the achievable rate.

*Theorem 3.4:* The FCF achievable rate for the DMC case is given as

$$R_{\text{FCF}} \leq \max \lambda I\left(X; \hat{Y}_1, \hat{Y}_2\right) \tag{27}$$

$$\text{s. t.} \quad C_{i,3-i} \geq \lambda I\left(\hat{Y}_i; Y_i\right) - \lambda I\left(\hat{Y}_i; Y_{3-i}\right), \ i = 1, 2, \tag{28}$$

$$\lambda I\left(\hat{Y}_1, \hat{Y}_2; Y_1, Y_2\right) \leq \overline{\lambda} I\left(X_r; Y\right), \tag{29}$$

over the distribution $p(x)p(y_1, y_2)p(\hat{y}_1|y_1)p(\hat{y}_2|y_2)p(x_r)p(y|x_r)$.

*Proof:* See Appendix B. ■

**Gaussian Case:** We choose the distributions of $X$ and $X_r$ as $\mathcal{CN}(0, P_S)$ and $\mathcal{CN}(0, P_r)$, respectively. Furthermore, the compressions at the relays are according to $\hat{Y}_i = Y_i + \hat{N}_i$, $i = 1, 2$.

*Corollary 3.4:* The FCF achievable rate for the Gaussian case is given as

$$R_{\text{FCF}} \leq \max_{\lambda, \sigma_1^2, \sigma_2^2} \lambda \log\left(1 + \frac{\gamma_1}{1+\sigma_1^2} + \frac{\gamma_2}{1+\sigma_2^2}\right) \tag{30}$$

$$\text{s. t.} \quad \sigma_i^2 \geq \frac{1+\gamma_1+\gamma_2}{\left(\gamma_{3-i}+1\right)\left(2^{C_{i,3-i}/\lambda}-1\right)}, \ i = 1, 2, \tag{31}$$

$$\lambda \log\left(1 + \frac{1+\gamma_1}{\sigma_1^2} + \frac{1+\gamma_2}{\sigma_2^2} + \frac{1+\gamma_1+\gamma_2}{\sigma_1^2\sigma_2^2}\right) \leq \overline{\lambda} \log\left(1 + \widetilde{\gamma}_1 + \widetilde{\gamma}_2 + 2\sqrt{\widetilde{\gamma}_1\widetilde{\gamma}_2}\right). \tag{32}$$

*Remark 3.9:* It can be checked that when $C_{i,3-i} = 0$, $R_{\text{FCF}} = 0$ for any channel parameters. This suggests that the FCF scheme is worse than the traditional CF scheme when $C_{i,3-i}$ is relatively small. In this case, we should not use conferencing to obtain full cooperation in the second hop, and the PCF scheme should be adopted instead. Denote the optimal solution for the CF rate (by Theorem 5.8 in [6]) as $\left(\overline{\sigma}_1^2, \overline{\sigma}_2^2, \overline{\lambda}\right)$, and it is easy to check that this solution also satisfies the constraint in (32). Thus, the threshold $\overline{C}_{i,3-i}$, below which the FCF scheme performs worse than the CF scheme, is obtained when the equality in (31) is achieved, i.e.,

$$\overline{C}_{i,3-i} = \overline{\lambda} \log\left(1 + \frac{1+\gamma_1+\gamma_2}{\overline{\sigma_i}^2\left(\gamma_{3-i}+1\right)}\right). \tag{33}$$

*Remark 3.10:* For any given finite $\gamma_i$ and $C_{i,3-i}$, when $\widetilde{\gamma}_i$ goes to infinity, the optimal $\lambda$ goes to 1. However, the compression noise power $\sigma_i^2$ cannot scale to 0 due to the constraints in (31), which means that the asymptotic capacity upper bound cannot be achieved.

*Remark 3.11:* For any given finite $\widetilde{\gamma}_i$ and $C_{i,3-i}$, when $\gamma_i \to \infty$ (assuming that $\gamma_1$ and $\gamma_2$ are on the same order), we choose $\lambda = \frac{\log\left(\widetilde{\gamma}_1+\widetilde{\gamma}_2+2\sqrt{\widetilde{\gamma}_1\widetilde{\gamma}_2}\right)}{\log(1+\gamma_1+\gamma_2)} \to 0$, while $\sigma_i^2$ scales on the order of $\frac{1}{\gamma_i}$ according to (31). For (32), it is easy to check that the left-hand side of the inequality is equal to the right-hand side asymptotically. Therefore, we conclude that *the FCF scheme asymptotically achieves the capacity upper bound as* $\gamma_i \to \infty$.

*Remark 3.12:* For any given finite $\gamma_i$ and $\widetilde{\gamma}_i$, whether there exist a finite set of $C_{12}$ and $C_{21}$ to achieve the capacity upper bound is also determined by whether (25) can meet the upper bound or not. By the same argument as in Remark 3.7, we conclude that *for fixed channel coefficients, the FCF scheme cannot achieve the capacity upper bound even with infinite conferencing link rates*.

*3) CCF Achievable Rate:* In this scheme, each relay generates its own compression intended for the second hop based on two signals: the received signal from the source, and the compressed signal from the other relay through the conferencing link.

**DMC Case:** We have the following theorem regarding the achievable rate.

*Theorem 3.5:* As we use the conferencing links to help with compressing the received signal at the relays, the CCF achievable rate for the DMC case is given by

$$\begin{aligned} R_{\text{CCF}} &= \max \lambda I\left(X;\hat{Y}_1,\hat{Y}_2\right) \\ \text{s. t. } (7),\ & \lambda I\left(\hat{Y}_1;Y_1,\hat{Y}_{21}|\hat{Y}_2\right) \le \overline{\lambda} I\left(X_1;Y|X_2\right) \\ & \lambda I\left(\hat{Y}_2;Y_2,\hat{Y}_{12}|\hat{Y}_1\right) \le \overline{\lambda} I\left(X_2;Y|X_1\right) \\ & \lambda I\left(\hat{Y}_1,\hat{Y}_2;Y_1,Y_2,\hat{Y}_{12},\hat{Y}_{21}\right) \le \overline{\lambda} I\left(X_1,X_2;Y\right), \end{aligned} \tag{34}$$

over the distribution $p(x)p(y_1,y_2|x)p\left(\hat{y}_{12}|y_1\right)p\left(\hat{y}_{21}|y_2\right)p\left(\hat{y}_1|y_1,\hat{y}_{21}\right)p\left(\hat{y}_2|y_2,\hat{y}_{12}\right)p(x_1,x_2)p(y|x_1,x_2)$.

*Proof:* See Appendix C. ■

**Gaussian Case:** We choose the distributions of transmit signals over the conferencing links as $\hat{Y}_{12} = Y_1 + N_{12}$ and $\hat{Y}_{21} = Y_2 + N_{21}$, respectively, where $N_{12}$ and $N_{21}$ are independent zero mean CSCG random variable, with variances defined the same as in (12). For the relay signals to the destination, we choose $\hat{Y}_1 = aY_1 + b\hat{Y}_{21} + V_1$ and $\hat{Y}_2 = cY_2 + d\hat{Y}_{12} + V_2$, where $a$, $b$, $c$, and $d$ are some parameters, $V_1$ and $V_2$ are independent zero mean CSCG random variables with variances $\sigma_1^2$ and $\sigma_1^2$, respectively. Then, the achievable rate for the Gaussian case is given as

$$R_{\text{CCF}} = \max_{\lambda,a,b,c,d,\hat{\sigma}_1^2,\hat{\sigma}_2^2} \lambda \log\left(\frac{P_{\hat{Y}_1\hat{Y}_2}}{\hat{\sigma}_1^2\hat{\sigma}_2^2 - |ad^* + bc^*|^2}\right) \tag{35}$$

$$\text{s. t.} \quad \lambda \log\left(\frac{P_{\hat{Y}_1\hat{Y}_2}}{\sigma_1^2\left(|dh_1 + ch_2|^2 P + \hat{\sigma}_2^2\right)}\right) \le \overline{\lambda} \log\left(1 + \widetilde{\gamma}_1\right)$$

$$\lambda \log\left(\frac{P_{\hat{Y}_1\hat{Y}_2}}{\sigma_2^2\left(|ah_1 + bh_2|^2 P + \hat{\sigma}_1^2\right)}\right) \le \overline{\lambda} \log\left(1 + \widetilde{\gamma}_2\right)$$

$$\lambda \log\left(\frac{P_{\hat{Y}_1\hat{Y}_2}}{\sigma_1^2\sigma_2^2}\right) \le \overline{\lambda} \log\left(1 + \widetilde{\gamma}_1 + \widetilde{\gamma}_2\right),$$

where

$$P_{\hat{Y}_1\hat{Y}_2} = |ah_1 + bh_2|^2 P_S \hat{\sigma}_2^2 + |dh_1 + ch_2|^2 P_S \hat{\sigma}_1^2 + \hat{\sigma}_1^2\hat{\sigma}_2^2 - |ad^* + bc^*|^2 - 2\Re\left[(ah_1 + bh_2)(dh_1 + ch_2)^* (ad^* + bc^*) P_S\right], \tag{36}$$

$\hat{\sigma}_1^2 = |a|^2 + |b|^2\left(1 + \sigma_{21}^2\right) + \sigma_1^2$, and $\hat{\sigma}_2^2 = |c|^2 + |d|^2\left(1 + \sigma_{12}^2\right) + \sigma_2^2$. It is easy to check that the above objective function is not convex over $a$, $b$, $c$, and $d$ jointly. Since it is difficult to compute the maximum rate, we try to find a sub-optimal but much simpler solution, i.e., letting $a = d = h_1^*$ and $b = c = h_2^*$, which will be used for the simulations in Section IV.

*Remark 3.13:* Since the traditional CF scheme is just a special case of our setup, by letting $\hat{Y}_{i,3-i}$ be a constant, the CCF achievable rate for the DMC case is the same as the case without conferencing [6]. Hence, with our setup, we conclude that the CCF rate is the same as or higher than the traditional CF rate. However, since only the sub-optimal solution for the combining problem at the relay is adopted, the CCF scheme may not perform better than the traditional CF scheme for the Gaussian case, and this will be shown in Section IV.

*Remark 3.14:* Consider another case when $C_{12}$ and $C_{21}$ go to infinity, while $\gamma_i$ and $\widetilde{\gamma}_i$ are finite. In this case, both of the relays could know $y_1$ and $y_2$, which corresponds to the perfect cooperation case. Then, the diamond relay channel becomes a two hop degraded relay channel. By the results of [1], we know that the CF scheme is strictly suboptimal, and there is a gap to the capacity upper bound in general. Therefore, we conclude that when the channel gains are fixed, even if $C_{i,3-i}$ goes to infinity, the CF scheme cannot achieve the capacity upper bound.

### D. AF Achievable Rate

In this subsection, to make the AF relaying scheme meaningful, we further assume that the conferencing links are Gaussian channels, which also use AF as the conferencing scheme. With-out loss of generality, we assume that the input of the conferencing link is $x_{i,3-i} = y_i = h_i x + n_i$. Furthermore, we assume that the link gain of each conferencing link equals to 1, and the conferencing link output in the $i$-th relay is given as

$$y_{3-i,i} = x_{3-i,i} + n_{3-i,i}, \tag{37}$$

where $n_{3-i,i}$ is CSCG noise with distribution $\mathcal{CN}\left(0, \sigma^2_{3-i,i}\right)$. Based on the conferencing link rate constraints, the variance of $n_{3-i,i}$ is given as $\sigma^2_{3-i,i} \geq \frac{\gamma_{3-i}+1}{2^{C_{3-i,i}/2}-1}$. Obviously, when the equality holds, the AF scheme performs the best. Thus, we let

$$\sigma^2_{3-i,i} = \frac{\gamma_{3-i}+1}{2^{C_{3-i,i}/2}-1}. \tag{38}$$

After the conferencing, the relays combine the two received signals from the source node and the other relay, which leads to

$$x_i = a_{ii} y_i + a_{3-i,i} y_{3-i,i}, \tag{39}$$

where $a_{ii}$ and $a_{3-i,i}$ are some complex parameters, and satisfy the following power constraints

$$\mathbb{E}\left(x_i^2\right) = |a_{ii}|^2 \left(|h_i|^2 P_S + 1\right) + |a_{3-i,i}|^2 \left(|h_{3-i}|^2 P_S + 1 + \sigma^2_{3-i,i}\right) \leq P_R. \tag{40}$$

Therefore, the received signal at the destination is given as

$$\begin{aligned} y &= g_1 x_1 + g_2 x_2 + n \\ &= \left(a_{11} h_1 g_1 + a_{12} h_1 g_2 + a_{21} h_2 g_1 + a_{22} h_2 g_2\right) x \\ &\quad + \left(a_{11} g_1 + a_{12} g_2\right) n_1 + \left(a_{21} g_1 + a_{22} g_2\right) n_2 + a_{21} g_1 n_{21} + a_{12} g_2 n_{12} + n, \end{aligned}$$

and the achievable rate of the AF scheme is given as

$$R_{\text{AF}} = \frac{1}{2} \log\left(1 + \gamma_{\text{AF}}\right), \tag{41}$$

where $\gamma_{\text{AF}}$ is the received SNR at the destination, given as

$$\gamma_{\text{AF}} = \frac{|a_{11} h_1 g_1 + a_{12} h_1 g_2 + a_{21} h_2 g_1 + a_{22} h_2 g_2|^2 P_S}{|a_{11} g_1 + a_{12} g_2|^2 + |a_{21} g_1 + a_{22} g_2|^2 + |a_{21} g_1|^2 \sigma^2_{21} + |a_{12} g_2|^2 \sigma^2_{12} + 1}. \tag{42}$$

We now rewrite (42) to a matrix form, and maximize it to obtain the maximum AF rate defined in (41). Thus, we have the following optimization problem

$$\begin{aligned} \max \quad & \frac{\mathbf{a}^H \mathbf{R} \mathbf{a}}{\mathbf{a}^H \mathbf{Q} \mathbf{a} + 1} \\ \text{s. t.} \quad & (40), \end{aligned} \tag{43}$$

where $\mathbf{a} = [a_{11}, a_{12}, a_{21}, a_{22}]^T$, $\mathbf{b} = [h_1^* g_1^*, h_1^* g_2^*, h_2^* g_1^*, h_2^* g_2^*]^T$, and the matrices $\mathbf{R} = \mathbf{b}\mathbf{b}^H$,

$$\mathbf{Q} = \begin{bmatrix} |g_1|^2 & g_1^* g_2 & 0 & 0 \\ g_1 g_2^* & |g_2|^2 (1 + \sigma_{12}^2) & 0 & 0 \\ 0 & 0 & |g_1|^2 (1 + \sigma_{21}^2) & g_1^* g_2 \\ 0 & 0 & g_1 g_2^* & |g_2|^2 \end{bmatrix}. \tag{44}$$

From (42), we know that $\mathbf{R}$ and $\mathbf{Q}$ are positive semidefinite. By a similar argument as in [22], this problem can be shown equivalent to

$$\max_{\mathbf{A}, t} \quad t \tag{45}$$

$$\text{s. t.} \quad \text{Tr}\left(\mathbf{A}\left(\mathbf{R} - t\mathbf{Q}\right)\right) \geq t, \ (40), \ \text{Rank}(\mathbf{A}) = 1, \mathbf{A} \succeq 0,$$

where $\mathbf{A} = \mathbf{a}\mathbf{a}^H$. Using semidefinite relaxation [22], we aim to solve the following optimization problem:

$$\max_{\mathbf{A}, t} \quad t \tag{46}$$

$$\text{s. t.} \quad \text{Tr}\left(\mathbf{A}\left(\mathbf{R} - t\mathbf{Q}\right)\right) \geq t, \ (40), \ \mathbf{A} \succeq 0.$$

*Remark 3.15:* This optimization problem can be efficiently solved by bisection search over $t$; and for each $t$, the remaining problem is a convex feasibility problem, which can be efficiently solved using existing numerical tools, e.g., CVX [23]. However, the final solution may not be rank-1 to satisfy the constraint in (45); so we use the following randomization technique [22] to provide an approximate solution to the original rank-1 problem in (45): Denote the solution of problem (46) as $\mathbf{A}^*$, with its eigenvalue decomposition $\mathbf{A}^* = \mathbf{U}\mathbf{D}\mathbf{U}^H$; we choose $\mathbf{a} = \mathbf{U}\mathbf{D}^{1/2}\mathbf{v}$, where $\mathbf{v}$ is a vector of zero-mean unit-variance i.i.d. Gaussian random variables. We then scale $\mathbf{a}$ to make the power constraints (40) satisfied [24].

*Remark 3.16:* If a rank-one optimal solution for (46) can be found, our AF rate will be higher than the AF rate without conferencing, i.e., the case $C_{i,3-i} = 0$. This is due to the facts that the traditional AF relaying optimization problem is a special case of (43) with $a_{12} = a_{21} = 0$. However, sometimes we may not obtain the exact optimal solution of rank-one for (46), such that there is a gap to the optimal value with the solution from the randomization method [24]. For these cases, our proposed AF scheme may not be better than the case without conferencing. By the results shown in Section IV, we observe that for small conferencing link rates, our scheme performs worse than the traditional AF scheme without conferencing; but the reverse is true for large $C_{i,3-i}$ cases.

*Remark 3.17:* It is easy to check that when $\gamma_i$ goes to infinity, the AF scheme can achieve one-half of the capacity upper bound, which is due to the half-duplex constraint. On the other hand, if both $\gamma_i$ and $\widetilde{\gamma}_i$ are finite, the upper bound is not achievable even with infinite conferencing link capacity.

## IV. Numerical Results

In this section, we present some numerical results to compare the performance among the proposed coding schemes. For simplicity, we only consider the symmetric case, i.e., $|h_1| = |h_2|$, $|g_1| = |g_2|$, and $C_{12} = C_{21} = C$. Set the locations of the source node, the destination node, and the relays as $s_0 = (-1, 0)$, $s_3 = (0, 1)$, $s_1 = (d, -\sqrt{1-d^2})$, and $s_2 = (d, +\sqrt{1-d^2})$, respectively, where $d \in (-1, 1)$. Furthermore, we assume that the link gains satisfy $|h_i| = \frac{1}{|s_0 - s_i|}$ and $|g_i| = \frac{1}{|s_3 - s_i|}$, $i = 1, 2$. For the phases of $h_i$ and $g_i$, we assume that they are uniform random variables over $[0, 2\pi]$.

In Fig. 3, we compare the performance of the proposed schemes with the conferencing link rate $C = 0.5$ bit/s/Hz. We observe that when $d$ goes to $-1$, i.e., when the relays get close to the source node, the DF scheme asymptotically achieves the capacity upper bound, so do the FCF and PCF schemes. Moreover, all three CF schemes outperform the AF scheme, but they are worse than the DF scheme. As $d$ goes to 1, i.e., when the relays get close to the destination, we observe that the PCF scheme achieve the capacity upper bound asymptotically, while the DF, AF, and FCF schemes are strictly suboptimal. For the case when $d$ is around 0, the DF scheme performs the best among all the achievable schemes, and the performances of the others are almost the same.

In Fig. 4(a) and Fig. 4(b), with different channel gains, we compare the performances of the coding schemes as the conferencing link rate increases. We consider two typical setups: the BC channel gains are larger than those of the MAC channel for Fig. 4(a), and the reverse case for Fig. Fig. 4(b). Overall, we observe that for each relaying scheme, there is an asymptotic performance limitation as the conferencing link rate increases.

Note that when $C = 0$, the proposed DF and PCF schemes are equivalent to the traditional DF and CF schemes. From these two subfigures, we observe that conferencing can strictly increase the DF and CF achievable rates using the proposed DF and PCF schemes, respectively. However, for the AF, CCF, and FCF schemes, they cannot guarantee to increase the AF and CF rates as we discussed before, respectively, especially when $C$ is small.

For both cases shown in Fig. 4(a) and Fig. 4(b), the DF scheme gets close to the capacity upper bound when $C$ is large enough: For the good BC channel case, we need $C \geq 2$ bits/s/Hz, and for the good MAC channel case, we need $C \geq 4$ bits/s/Hz. For the PCF and FCF schemes, we observe that as $C$ becomes large, they have the same performance; when $C$ is very close to 0, the PCF scheme always performs better; for small $C$ but not close to 0, the FCF scheme performs better in the good BC channel case, and the reverse is true for the good MAC channel case. In the high conferencing rate regime, the CCF scheme performs better than the other two CF schemes for the good MAC channel case, and the reverse is true for the good BC channel case.

## V. Conclusion

In this paper, we discussed the capacity upper bound and the achievable rates of the diamond relay channel with conferencing links. For the DF scheme, we derived the achievable rate by sending a common message and two private messages. We proved that for the DMC case, the DF scheme can achieve the capacity upper bound with finite conferencing link rates, which is not true for the Gaussian case. Moreover, the DF scheme is asymptotically optimal when the link SNRs of the first hop go to infinity. We developed three new coding schemes based on CF and used the conferencing links to exchange certain compressed information between the relays. The achievable rates were computed for both the DMC and Gaussian cases, and the capacity-achieving cases were discussed. For the AF scheme, we discussed the optimal combining problem between the signals from the source and the conferencing link at the relays, and use semidefinite relaxation and bisection search to efficiently obtain a sub-optimal solution.

## Appendix A
## Proof of Lemma 3.1

Fix the distribution $p(u_0)p(u_1|u_0)p(u_2|u_0)p(y_1, y_2|x)\ p(\hat{y}_1|y_1)p(\hat{y}_2|y_2)$ and the function $x(u_0, u_1, u_2)$.

**Codebook Generation:** In the source, generate $2^{nR_0}$ i.i.d. sequences $\boldsymbol{u}_0\left(w_0\right)$, $w_0 \in \left[1 : 2^{nR_0}\right]$, according to the distribution $\prod_{j=1}^{\lambda n} p\left(u_{0,j}\right)$. For each $\boldsymbol{u}_0\left(w_0\right)$, generate $2^{nR_i}$ i.i.d. sub-codebooks $\boldsymbol{Q}_i(w_0, w_i)$, $w_i \in \left[1 : 2^{nR_i}\right]$, where each sub-codebook contains $2^{n\left(\widetilde{R}_i - R_i\right)}$ i.i.d. sequences $\boldsymbol{u}_i(w_0, l_i)$, $l_i \in \left[(w_i - 1)\, 2^{n\left(\widetilde{R}_i - R_i\right)} + 1 : w_i 2^{n\left(\widetilde{R}_i - R_i\right)}\right]$, according to $\prod_{j=1}^{\lambda n} p\left(u_{i,j} | u_{0,j}(w_0)\right)$. For each triple $(w_0, w_1, w_2)$, define the set

$$\begin{aligned}\boldsymbol{Q}(w_0, w_1, w_2) = \{&(u_1(w_0, l_1), u_2(w_0, l_2)) \in \boldsymbol{Q}_1(w_0, w_1) \times \boldsymbol{Q}_2(w_0, w_2) : \\ &(u_0(w_0), u_1(w_0, l_1), u_2(w_0, l_2)) \in A_{\epsilon'}^n\}.\end{aligned}$$

**Conferencing function generation:** Generate $2^{nR_i'}$ i.i.d. sequences $\hat{\boldsymbol{y}}_i(k_i)$, $k_i \in \left[1 : 2^{nR_i'}\right]$, according to $\prod_{j=1}^{\lambda n} p\left(\hat{y}_{i,j}\right)$, where $p_{\hat{Y}_i}\left(\hat{y}_{i,j}\right) = \sum_{\mathcal{X}, \mathcal{Y}_1, \mathcal{Y}_2} p\left(\hat{y}_i | y_i\right) p\left(y_1, y_2 | x\right) p(x)$ and $p(x) = \sum_{\mathcal{U}_1, \mathcal{U}_2} p(u_1, u_2, x)$. Randomly and uniformly partition the index set $\left[1 : 2^{nR_i'}\right]$ into $2^{nC_{i,3-i}}$ binnings $\boldsymbol{S}_i(s_i)$, $s_i \in \left[1 : 2^{nC_{i,3-i}}\right]$.

**Encoding and Decoding:** In the source, for each triple $(w_0, w_1, w_2)$, pick one sequence pair $(\boldsymbol{u}_1(w_0, l_1), \boldsymbol{u}_2(w_0, l_2)) \in \boldsymbol{Q}(w_0, w_1, w_2)$, and generate a codeword $\boldsymbol{x}(w_0, w_1, w_2)$ according to $\prod_{i=1}^{\lambda n} p\left(x_i | u_1(w_0, l_1), u_2(w_0, l_2)\right)$; if no such pair exists, declare an error. This operation can be done reliably if [25]

$$\left(\widetilde{R}_1 - R_1\right) + \left(\widetilde{R}_2 - R_2\right) \geq \lambda I\left(U_1; U_2 | U_0\right). \tag{47}$$

In the $i$-th relay, upon receiving $\boldsymbol{y}_i$, it tries to find a $\hat{\boldsymbol{y}}_i(k_i)$ such that $(\boldsymbol{y}_i, \hat{\boldsymbol{y}}_i(k_i)) \in A_\epsilon^n$, and this can be done reliably as $n$ goes to infinity, if

$$R_i' \geq \lambda I\left(\hat{Y}_i; Y_i\right). \tag{48}$$

Then, the $i$-th relay finds the corresponding binning index number $s_i$, where $k_i \in \boldsymbol{S}_i(s_i)$, and sends it to the other relay through the conferencing link.

After receiving the conferencing message from its counterpart, the $i$-th relay first tries to find the unique $\hat{k}_{3-i}$ such that $\left(\hat{\boldsymbol{y}}_{3-i}(\hat{k}_{3-i}), \boldsymbol{y}_i\right) \in A_\epsilon^n$ with $\hat{k}_{3-i} \in \boldsymbol{S}_{3-i}(s_{3-i})$. This can be done reliably if

$$R_{3-i}' \leq \lambda I\left(\hat{Y}_{3-i}; Y_i\right) + C_{3-i,i}. \tag{49}$$

From (48) and (49), we obtain

$$C_{i,3-i} \geq \lambda I\left(\hat{Y}_i; Y_i\right) - \lambda I\left(\hat{Y}_i; Y_{3-i}\right). \tag{50}$$

Then, the $i$-th relay finds a unique pair $(\hat{w}_0, \hat{w}_i)$ satisfying $\left(\boldsymbol{u}_0(\hat{w}_0), \boldsymbol{u}_i(\hat{w}_0, \hat{l}_i), \hat{\boldsymbol{y}}_{3-i}(\hat{k}_{3-i}), \boldsymbol{y}_i\right) \in A_\epsilon^n$, and this can be done reliably if

$$\begin{cases} \widetilde{R}_i \leq \lambda I\left(U_i; \hat{Y}_{3-i}, Y_i | U_0\right) \\ R_0 + \widetilde{R}_i \leq \lambda I\left(U_0, U_i; \hat{Y}_{3-i}, Y_i\right). \end{cases} \tag{51}$$

From (47), (50), and (51), we obtain the rate region of the general broadcast channel with common message and conferencing as follows:

$$R'_{\text{BC}} = \bigcup_{p(u_0)p(u_1|u_0)p(u_2|u_0)p(x|u_1,u_2)p(y_1,y_2|x)p(\hat{y}_1|y_1)p(\hat{y}_2|y_2)} \left\{ \begin{array}{l} (R_0, R_1, R_2): \\ 0 \leq R_0, 0 \leq R_1 \leq \widetilde{R}_1, 0 \leq R_2 \leq \widetilde{R}_2, \\ \widetilde{R}_1 \leq \lambda I\left(U_1; \hat{Y}_2, Y_1 | U_0\right), \\ R_0 + \widetilde{R}_1 \leq \lambda I\left(U_0, U_1; \hat{Y}_2, Y_1\right), \\ \widetilde{R}_2 \leq \lambda I\left(U_2; \hat{Y}_1, Y_2 | U_0\right), \\ R_0 + \widetilde{R}_2 \leq \lambda I\left(U_0, U_2; \hat{Y}_1, Y_2\right), \\ \left(\widetilde{R}_1 - R_1\right) + \left(\widetilde{R}_2 - R_2\right) \geq \lambda I\left(U_1; U_2 | U_0\right), \\ \text{subject to: } (50). \end{array} \right\}. \tag{52}$$

Thus, the rate region $R_{\text{BC}}$ is obtained from $R'_{\text{BC}}$ using the Fourier-Motzkin elimination [26] to eliminate $\widetilde{R}_i$, $i = 1, 2$.

## APPENDIX B
## PROOF OF THEOREM 3.4

Fix the distribution as given in the theorem.

**Codebook generation:** Generate $2^{nR}$ i.i.d. sequences $\boldsymbol{x}(w)$, $w \in \left[1 : 2^{nR}\right]$, according to $\prod_{i=1}^n p(x_i)$. Generate $2^{n\hat{R}_i}$, $i = 1, 2$, i.i.d. sequences $\hat{\boldsymbol{y}}_i(w_i)$, $w_i \in \left[1 : 2^{n\hat{R}_i}\right]$, according to the distribution $p(\hat{y}_i) = \int p(x)p(y_i|x)p(\hat{y}_i|y_i)dxdy_i$. Randomly and uniformly partition the set $\left[1 : 2^{n\hat{R}_i}\right]$ into $2^{nR_i}$ binnings $\boldsymbol{S}_i(s_i)$, $s_i \in \left[1 : 2^{nR_i}\right]$. Randomly and uniformly partition the set $\left[1 : 2^{nR_i}\right]$ into $2^{nC_{i,3-i}}$ binnings $\boldsymbol{M}_i(m_i)$. Generate $2^{n(R_1+R_2)}$ i.i.d. sequences $\boldsymbol{x}_r(s_1, s_2)$, according to $p(x_r)$.

**Encoding and decoding:** At the source, it transmits $\boldsymbol{x}(w)$. At the $i$-th relay, $i = 1, 2$, it finds a $\hat{\boldsymbol{y}}_i(w_i)$ such that $(\hat{\boldsymbol{y}}_i(w_i), \boldsymbol{y}_i) \in A_\epsilon^n$, and this can be done reliably if $\hat{R}_i \geq \lambda I\left(\hat{Y}_i; Y_i\right)$.

Then, at the $i$-th relay, it finds the conferencing binning index $m_i$, and sends it to the other relay through the conferencing link. Upon receiving $m_{3-i}$, the $i$-th relay decodes $\hat{\boldsymbol{y}}_{3-i}(w_{3-i})$ such that $\left(\hat{\boldsymbol{y}}_{3-i}(\hat{k}_{3-i}), \boldsymbol{y}_i\right) \in A_\epsilon^n$ with $\hat{k}_{3-i} \in \boldsymbol{M}_{3-i}(m_{3-i})$. This can be done reliably if $R'_{3-i} \leq \lambda I\left(\hat{Y}_{3-i}; Y_i\right) + C_{3-i,i}$. Thus, we satisfy the constraints in (28). Then, the $i$-th relay knows the binning index pair $(s_1, s_2)$, and transmits $\boldsymbol{x}_r(s_1, s_2)$.

At the destination, it first decodes $(s_1, s_2)$, and we obtain $R_1 + R_2 \leq \overline{\lambda} I\left(X_r; Y\right)$. Then, the destination decodes $(\hat{\boldsymbol{y}}_1, \hat{\boldsymbol{y}}_2)$ and the original message $w$. By a similar argument as in Section VC of [6], we obtain (29).

## Appendix C
## Proof of Theorem 3.5

First fix the distribution as shown in the theorem.

**Codebook Generation:** Generate $\boldsymbol{x}(w)$ the same as those in Appendix B. Generate $2^{nR_{i,3-i}}$ i.i.d. sequences $\hat{\boldsymbol{y}}_{i,3-i}(k_i)$, according to $\prod_{j=1}^{\lambda n} p(\hat{y}_{i,3-i}^j)$ with $p\left(\hat{y}_{i,3-i}\right) = \int p\left(y_i\right) p\left(\hat{y}_{i,3-i}|y_i\right) dy_i$. Randomly and uniformly partition the set $\left[1 : 2^{nR_{i,3-i}}\right]$ into $2^{nC_{i,3-i}}$ bins $\boldsymbol{S}_{i,3-i}(s_{i,3-i})$; generate $2^{nR_{i0}}$ i.i.d. sequences $\hat{\boldsymbol{y}}_i(w_i)$, according to $\prod_{j=1}^{\lambda n} p\left(\hat{y}_{i,j}\right)$ with $p\left(\hat{y}_i\right) = \int p\left(\hat{y}_i|y_i, \hat{y}_{3-i,i}\right) p\left(y_i, \hat{y}_{3-i,i}\right) dy_i d\hat{y}_{3-i,i}$. Randomly and uniformly partition the set $\left[1 : 2^{nR_{i0}}\right]$ into $2^{nR_i}$ bins $\widetilde{\boldsymbol{S}}_i(\widetilde{s}_i)$; and generate $2^{nR_i}$ i.i.d. sequences $\boldsymbol{x}_i(\widetilde{s}_i)$, according to $p_{X_i}(x_i)$.

**Encoding and Decoding:** At the source, it transmits $\boldsymbol{x}(w)$; in the $i$-th relay, the conferencing scheme is the same as the DF scheme, which is omitted here; and we obtain (7). Based on $\boldsymbol{y}_i$ and $\hat{\boldsymbol{y}}_{3-i,i}$, the $i$-th relay find a $\hat{\boldsymbol{y}}_i\left(k_i\right)$ such that $\left(\hat{\boldsymbol{y}}_i\left(k_i\right), \hat{\boldsymbol{y}}_{3-i,i}\left(w_{3-i,i}\right), \boldsymbol{y}_i\right) \in A_\epsilon^n$, and this can be done reliably if $R_{i0} \geq I\left(\hat{Y}_i; \hat{Y}_{3-i,i}, Y_i\right)$. Then, the $i$-th relay obtains the binning index $\widetilde{s}_i$ and sends $\boldsymbol{x}_i(\widetilde{s}_i)$ to the destination.

In the destination, upon receiving $\boldsymbol{y}$, it first decodes the pair $(\widetilde{s}_1, \widetilde{s}_2)$, and the rate region $(R_1, R_2)$ is given by the MAC rate region as in [20], [25]. Then, the destination tries to decode $(\hat{\boldsymbol{y}}_1, \hat{\boldsymbol{y}}_2)$. Following a similar argument as in [6], [21], we have

$$\begin{cases} R_1 \geq \lambda I\left(\hat{Y}_1; Y_1, \hat{Y}_{21}|\hat{Y}_2\right) \\ R_2 \geq \lambda I\left(\hat{Y}_2; Y_2, \hat{Y}_{12}|\hat{Y}_1\right) \\ R_1 + R_2 \geq \lambda I\left(\hat{Y}_1, \hat{Y}_2; Y_1, Y_2, \hat{Y}_{12}, \hat{Y}_{21}\right) \end{cases}. \tag{53}$$

Finally, by finding a unique $\hat{w}$ such that $(\boldsymbol{x}(\hat{w}), \hat{\boldsymbol{y}}_1, \hat{\boldsymbol{y}}_2) \in A_\epsilon^n$, we obtain $R_{\text{CF}} = \lambda I\left(X; \hat{Y}_1, \hat{Y}_2\right)$. With the Fourier-Motzkin elimination [26], and the facts that $I\left(\hat{Y}_1, \hat{Y}_2; Y_1, Y_2, \hat{Y}_{12}, \hat{Y}_{21}\right) \geq I\left(\hat{Y}_1; Y_1, \hat{Y}_{21} | \hat{Y}_2\right) + I\left(\hat{Y}_2; Y_2, \hat{Y}_{12} | \hat{Y}_1\right)$ and $I\left(X_1, X_2; Y\right) \leq I\left(X_1; Y | X_2\right) + I\left(X_2; Y | X_1\right)$, the theorem is proved.

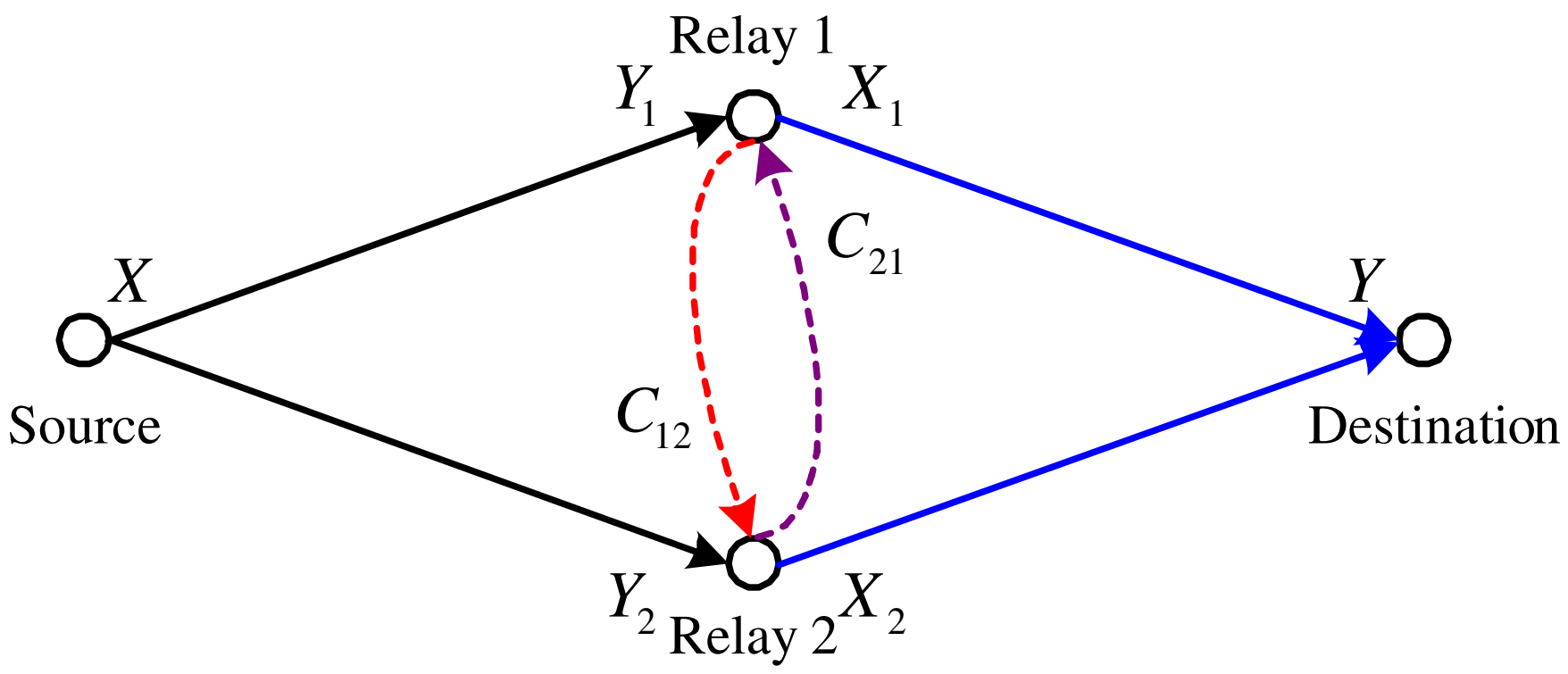

Fig. 1. Diamond relay channel with conferencing links.

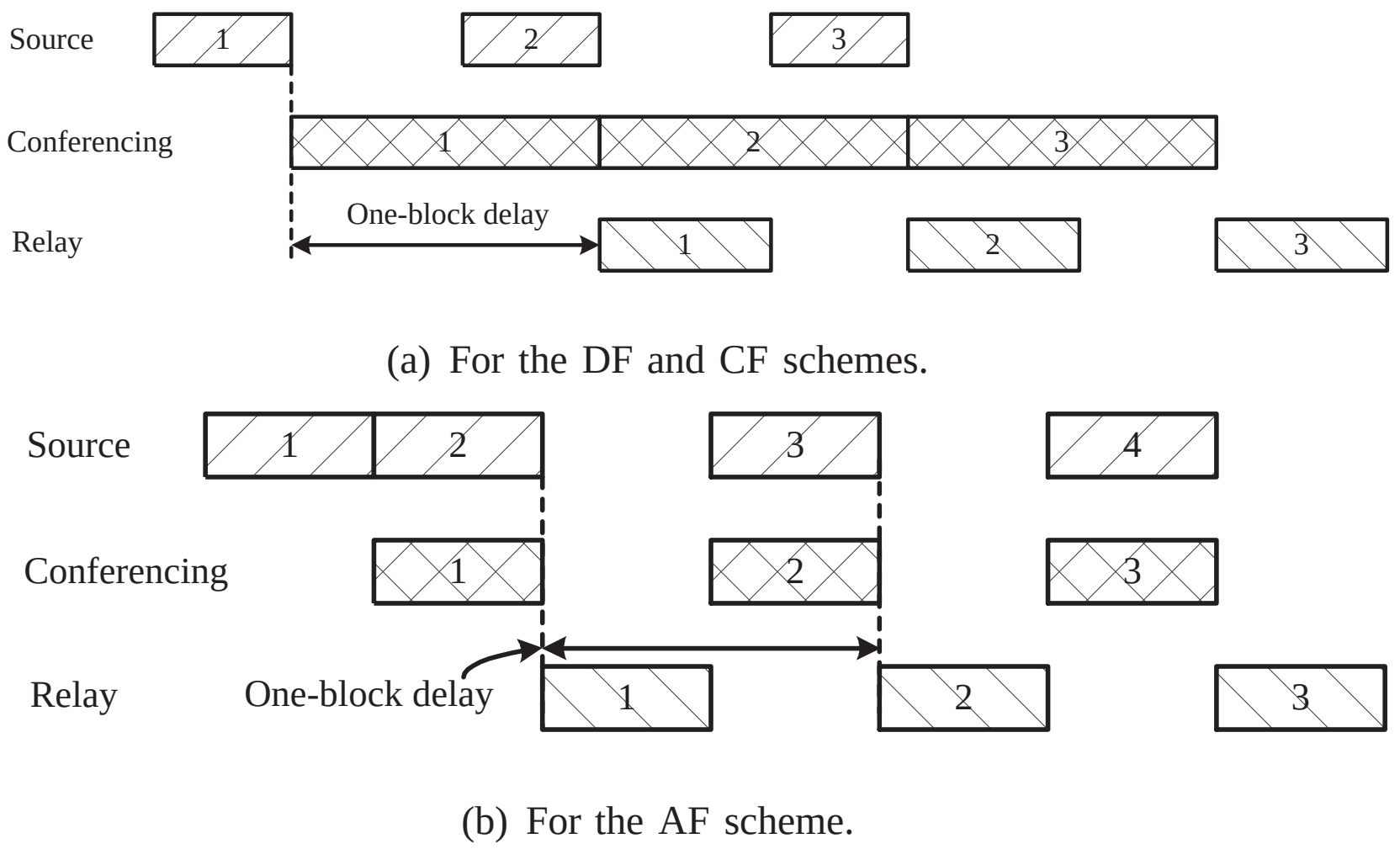

Fig. 2. Transmission scheduling scheme for the diamond relay channel with conferencing links.

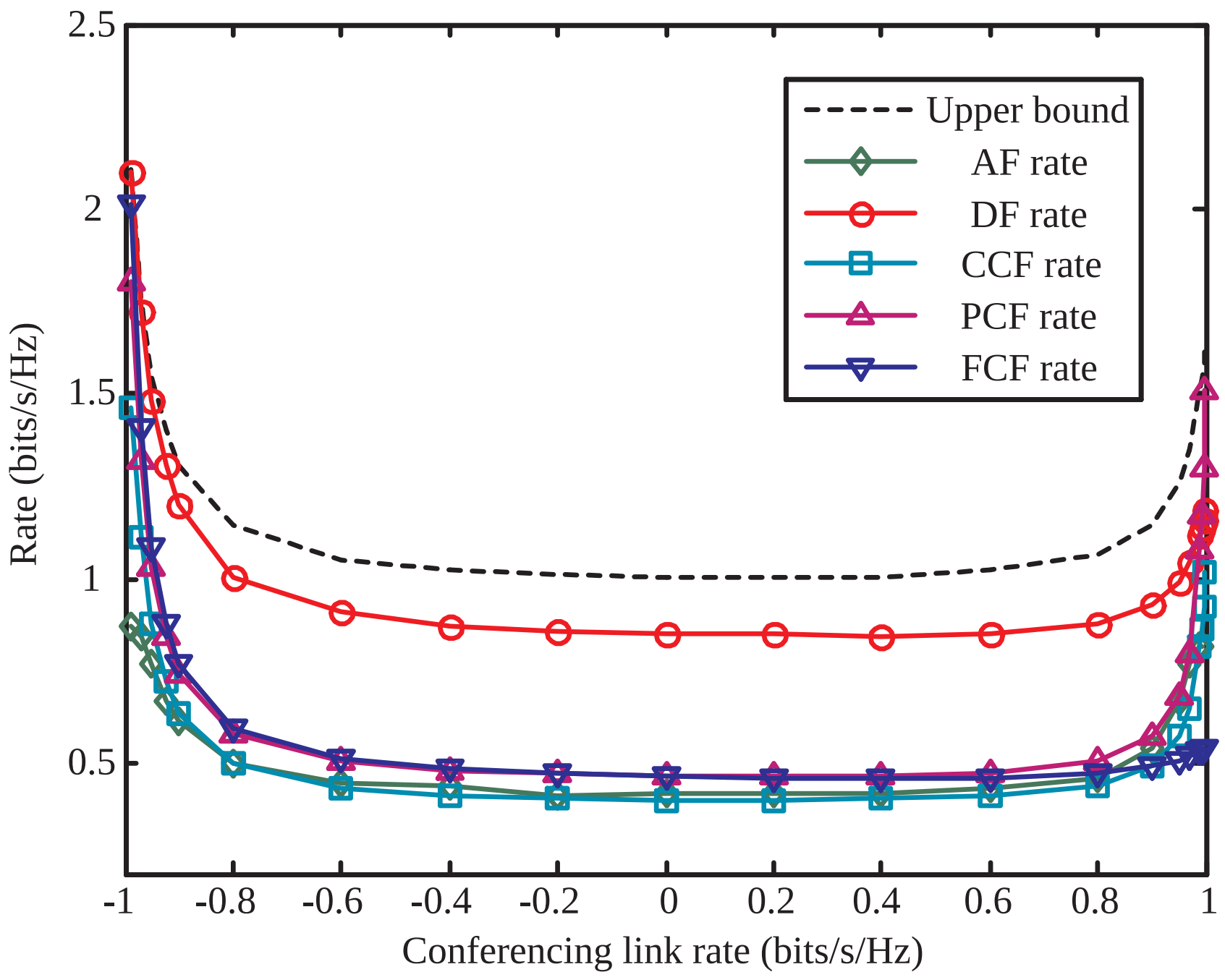

Fig. 3. The achievable rates and cut-set upper bound for symmetric link gain case, with $C = 0.5$.

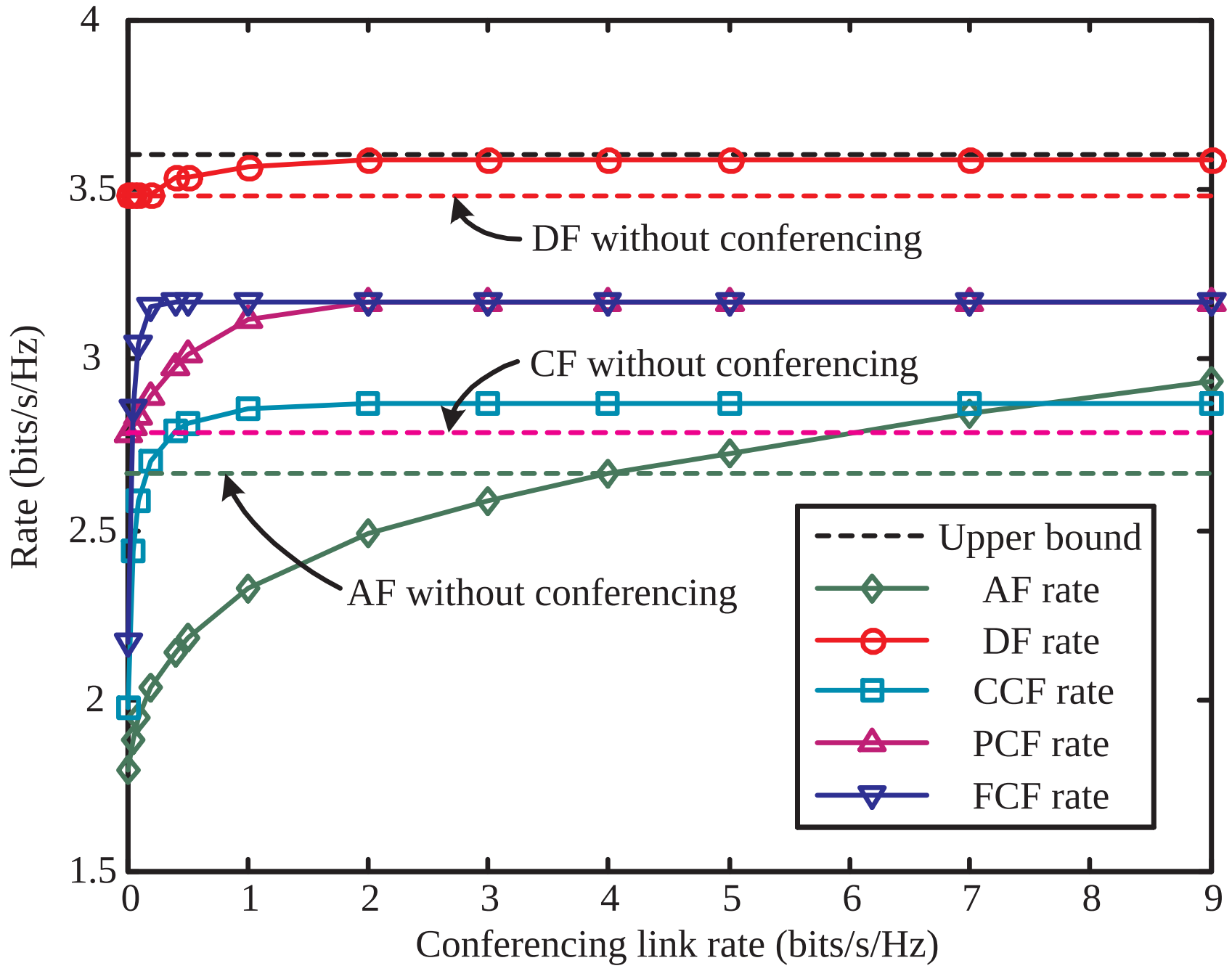

(a) Good BC vs. bad MAC, $\gamma_i = 30$ dB, $\widetilde{\gamma}_i = 10$ dB

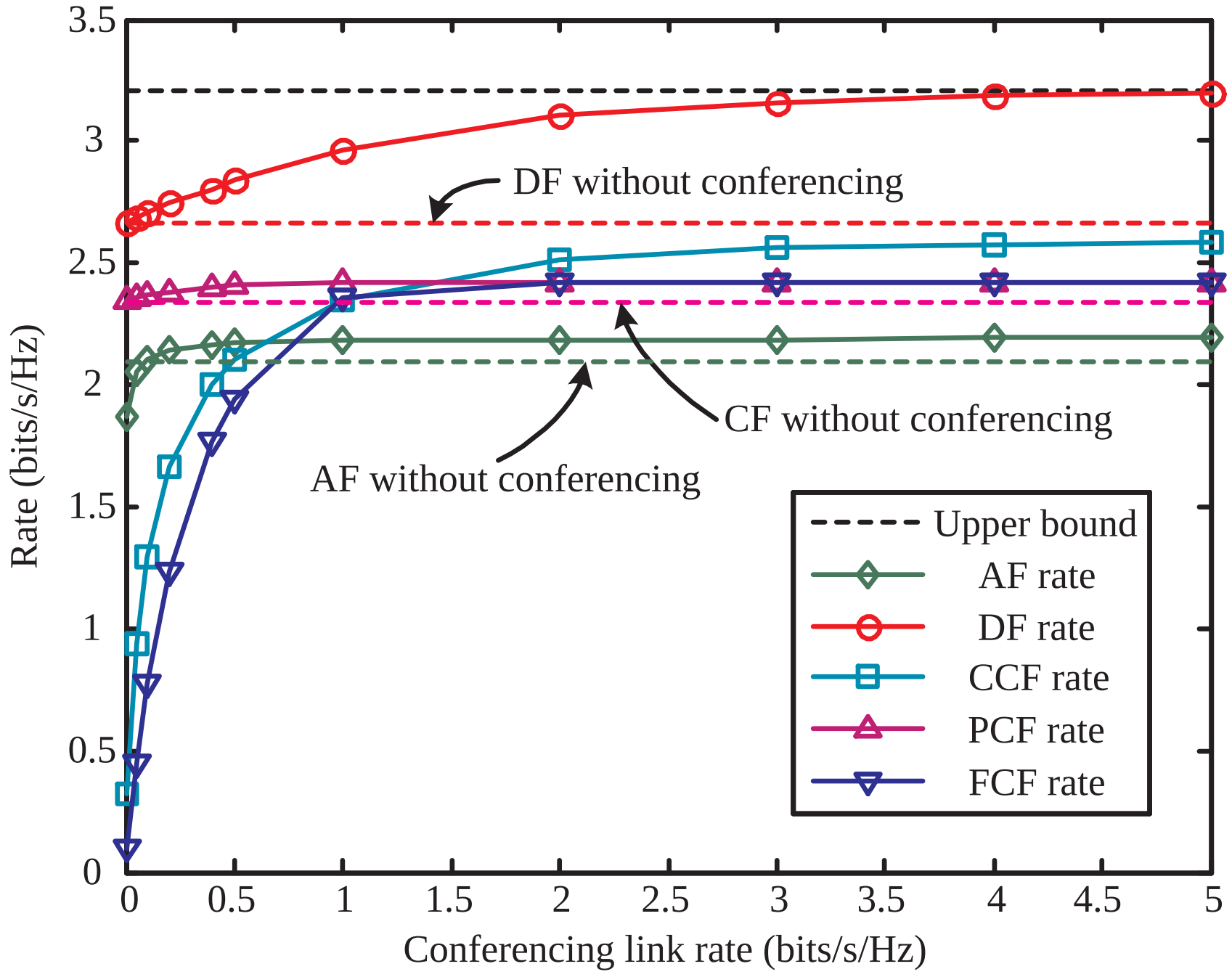

(b) Bad BC vs. good MAC, $\gamma_i = 10$ dB, $\widetilde{\gamma}_i = 30$ dB

Fig. 4. The achievable rates for different conferencing link rates, $P_S = P_R = 1$.